\documentclass[sigconf]{acmart}
\usepackage{graphicx}
\usepackage[ruled]{algorithm2e}
\usepackage{color}
\usepackage{enumitem}
\usepackage{float}
\usepackage{pifont}
\usepackage{makecell}

\AtBeginDocument{
  }

\setcopyright{acmlicensed}

\copyrightyear{2026}
\acmYear{2026}
\setcopyright{cc}
\setcctype{by}
\acmConference[KDD '26]{Proceedings of the 32nd ACM SIGKDD Conference on Knowledge Discovery and Data Mining V.2}{August 09--13, 2026}{Jeju Island, Republic of Korea}
\acmBooktitle{Proceedings of the 32nd ACM SIGKDD Conference on Knowledge Discovery and Data Mining V.2 (KDD '26), August 09--13, 2026, Jeju Island, Republic of Korea}
\acmDOI{10.1145/3770855.3818434}
\acmISBN{979-8-4007-2259-2/2026/08}

\begin{document}

\title{HMAF: A Hierarchical Multi-Slot GD-RTB Allocation Framework}

\author{Tianxing Bu}
\affiliation{%
  \institution{Meituan}
  \city{Beijing}
  \country{China}
}
\email{butianxing@meituan.com}

\author{Zhaoqi Zhang}
\authornote{Corresponding author.}
\affiliation{%
  \institution{Nanyang Technological University}
  \country{Singapore}
}
\email{zhaoqi001@e.ntu.edu.sg}

\author{Linyou Cai}
\affiliation{%
  \institution{Meituan}
  \city{Beijing}
  \country{China}
}
\email{cailinyou@meituan.com}

\author{Miao Xie}
\affiliation{%
  \institution{China Agricultural University}
  \city{Beijing}
  \country{China}
}
\email{xiemiao@cau.edu.cn}

\author{Shengri Xue}
\affiliation{%
  \institution{Meituan}
  \city{Beijing}
  \country{China}
}
\email{xueshengri@meituan.com}

\author{Tan Qu}
\affiliation{%
  \institution{Meituan}
  \city{Beijing}
  \country{China}
}
\email{qutan@meituan.com}

\author{Qianlong Xie}
\affiliation{%
  \institution{Meituan}
  \city{Beijing}
  \country{China}
}
\email{xieqianlong@meituan.com}

\author{Xingxing Wang}
\affiliation{%
  \institution{Meituan}
  \city{Beijing}
  \country{China}
}
\email{wangxingxing04@meituan.com}

\author{Siqiang Luo}
\affiliation{%
  \institution{Nanyang Technological University}
  \country{Singapore}
}
\email{siqiang.luo@ntu.edu.sg}

\author{Gao Cong}
\affiliation{%
  \institution{Nanyang Technological University}
  \country{Singapore}
}
\email{gaocong@ntu.edu.sg}

\renewcommand{\shortauthors}{Tianxing Bu et al.}

\begin{abstract}


In modern online advertising platforms, Guaranteed Delivery (GD) contracts coexist and bid with Real-Time Bidding (RTB) auctions. 
Recent approaches either decouple GD and RTB optimization or rely on heuristic priority rules, and thus fail to effectively balance short-term revenue maximization with long-term contract delivery under complex multi-slot delivery and impression constraints.
To address these challenges, we propose HMAF (Hierarchical Multi-Slot Allocation Framework), a unified framework designed to optimize impression allocation in GD-RTB advertising platforms. HMAF employs the \emph{Plan–Calibrate–Execute} paradigm as its core structure, and integrates offline constraint optimization with online decision-making, balancing offline GD resource planning, 
dynamically calibrating GD-RTB competitiveness, and making real-time listwise rank decisions across multi-slot environments. HMAF has been implemented in multiple marketing scenarios at Meituan, one of the world’s largest online food delivery platforms, leading to a 3.72\% increase on GD delivery rate and 1.59\% increase on total advertisement revenue.
\end{abstract}



\maketitle

\section{Introduction}


In large-scale online advertising ecosystems, especially multi-slot environments such as Online-to-Offline content feeds, impression allocation is governed by a hybrid marketplace where Guaranteed Delivery (GD) contracts~\cite{ghosh2009bidding, bhalgat2012online, hojjat2014delivering} coexist and bid with Real-Time Bidding (RTB) auctions~\cite{yuan2013real, zhang2014optimal, wang2015real}. GD contracts ensure long-term contract delivery by guaranteeing fixed impression volumes at pre-negotiated prices, while RTB auctions dynamically monetize remaining inventory through competitive bidding. Although revenue maximization is relatively straightforward in a pure RTB setting, hybrid markets introduce an inherent conflict between short-term eCPM-driven revenue efficiency and long-term delivery guarantees. This conflict is further amplified in multi-slot page view setting. 


Prior research has studied GD and RTB mechanisms largely in isolation or through loosely coupled pipelines. Early works~\cite{bharadwaj2012shale} on GD allocation formulated delivery as a constrained optimization problem, primarily focusing on offline planning to satisfy supply–demand balance. To cope with real-time traffic uncertainty, subsequent studies introduced feedback and control-theoretic approaches, including PID-based controllers~\cite{zhang2016feedback, chen2011real}, to dynamically regulate bidding prices or sampling rates. 
Recent studies consider GD and RTB competing in shared auctions.
CONFLUX~\cite{wang2022conflux} coordinates GD and RTB via a primal–dual control mechanism under delivery constraints,
while LIBRA~\cite{wang2024follow} embeds contract-aware signals into learning-based hybrid ranking. While existing systems mark an important step toward unified GD–RTB competition, they remain limited in their ability to address the full complexity of large-scale, multi-slot hybrid allocation.

Despite recent progress in GD-RTB allocation and delivery control, three fundamental challenges persist:
\textbf{(1) NP-Hardness under Fine-Grained Multi-Slot Constraints.} 
The hybrid allocation problem involves large-scale combinatorial optimization over millions of requests and thousands of contracts. Fine-grained supply-side constraints, such as Page View limits to mitigate ad fatigue, further complicate the problem and render global optimization NP-hard. 
\textbf{(2) eCPM-fulfillment Trade-off under Delivery Constraints.} 
Pursuing higher eCPM can improve short-term momonetization but often risks under-delivery when delivery constraints become binding, while overly aggressive fulfilment pacing may compromiseconomic efficiency. Effectively managing this constraint-aware trade-off is crucial for sustaining both revenue efficiency and delivery stability in large-scale hybrid advertising systems. 
\textbf{(3) Weak Supervision under Hybrid Ranking Objectives.} 
Unlike conventional CTR prediction tasks with explicit click labels, hybrid ranking lacks ground-truth supervision, defining an optimal ranking list that balances immediate revenue with long-term delivery guarantees. As a result, the system must learn from implicit and delayed feedback derived from aggregate delivery states rather than direct labels. 


\begin{figure}
    \centering
    \includegraphics[width=0.45\textwidth]{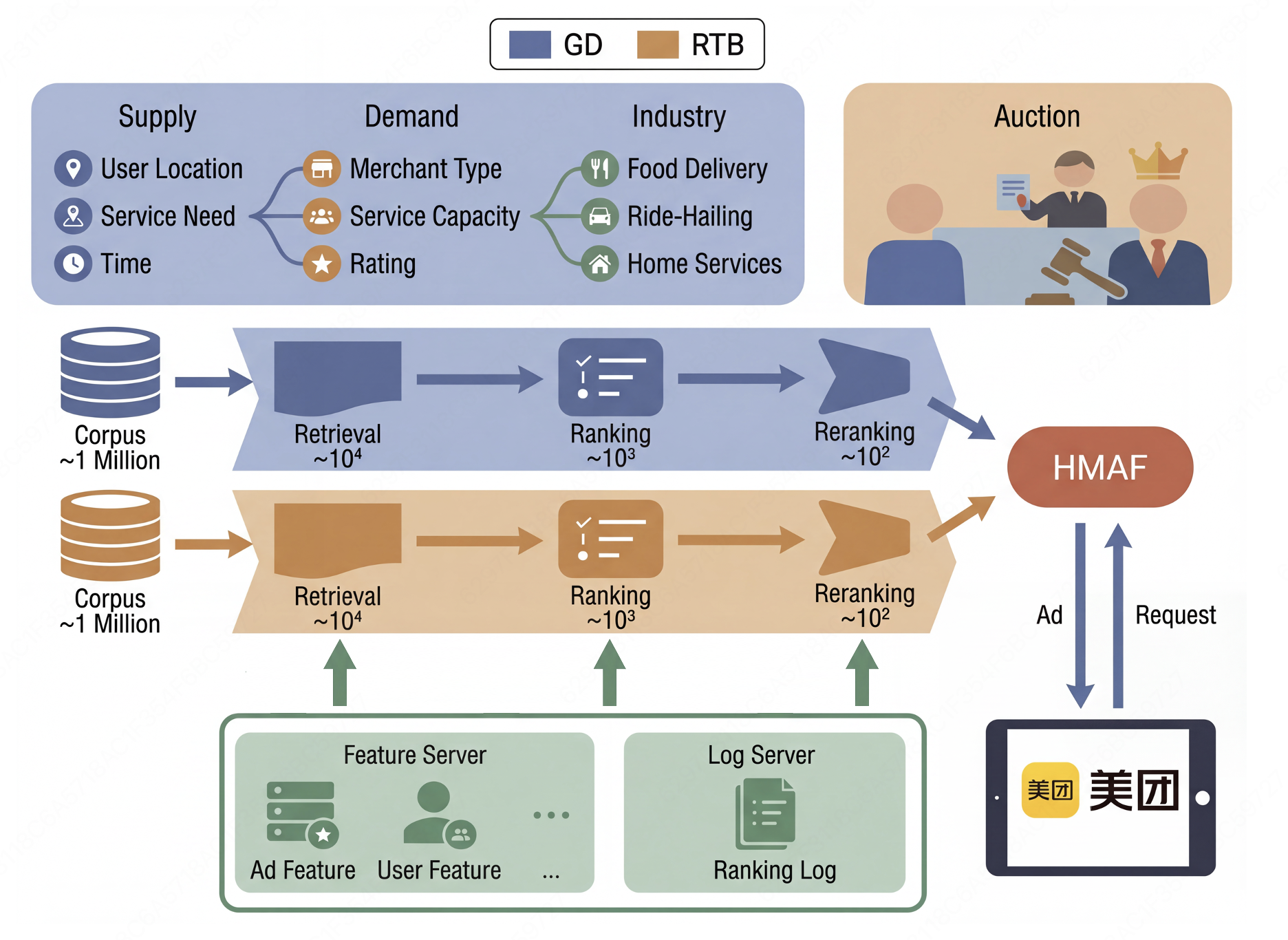}
    \caption{Flowchart of Meituan display advertising system}
    \label{figure: locate}
\end{figure}

To address the aforementioned challenges, we propose \textbf{HMAF}, a \textbf{H}ierarchical \textbf{M}ulti-Slot and \textbf{A}llowcation \textbf{F}ramework, which decomposes the intractable GD-RTB allocation problem into a hierarchical \emph{Plan-Calibrate-Execute} paradigm.
As illustrated in Figure~\ref{figure: locate}, HMAF is positioned at the confluence of two parallel pipelines, aggregating their outputs to construct a unified competitive stage for GD and RTB ads.
At the \emph{plan} stage, we formulate a \textbf{Page View-Constrained Optimization} problem and solve it offline to derive optimal dual variables. 
These dual variables act as shadow prices that encode the tightness of delivery constraints and the marginal value of page-view resources for individual contracts and slots, distilling long-horizon delivery pressure and supply-demand imbalance into compact control signals for downstream decision-making.
At the \emph{calibration} stage, we design a \textbf{Dual-Guided Unified Ranking} mechanism that injects offline dual variables into online scoring, balancing eCPM maximization with long-term fulfillment requirements under delivery and slot constraints.
At the \emph{execution} stage, we develop a \textbf{Generator-Evaluator} network for real-time GD-RTB ranking, where the dual variables are incorporated as global state features to guide listwise generation, allowing the model to trade off instantaneous RTB revenue against the shadow cost of potential GD violations and thereby transforming unsupervised ranking into a reward-guided decision process.

TThe main contributions of this paper are summarized as follows:
\begin{itemize}[leftmargin=*]
    \item 
    We propose HMAF, a unified GD-RTB allocation and ranking framework, which tightly couples offline constraint optimization with online neural decision making, providing a reliable architectural foundation for GD-RTB allocation in multi-slot advertising systems.
    \item 
    We introduce three tightly coupled core components, 
    a Page View-Constrained Optimization module, which provides a principled optimization backbone that captures fine-grained supply-side constraints and global resource scarcity;
    a Dual-Guided Unified Ranking module, which calibrates online candidate scoring using the offline dual variables, explicitly balancing eCPM maximization and contract fulfillment under delivery constraints;
    and a Generator-Evaluator module, which enables listwise GD-RTB ranking by jointly modeling candidate–position interactions and selecting revenue-optimal ranking lists under global guidance. 
    \item 
    We conduct large-scale comparative experiments and online
    A/B tests on Meituan’s advertising platform, HMAF achieved statistically significant improvements: 3.72\% increase on GD delivery rate and 1.59\% increase on total advertisement revenue
\end{itemize}

\begin{figure*}
    \centering
    \includegraphics[width=0.95\textwidth]{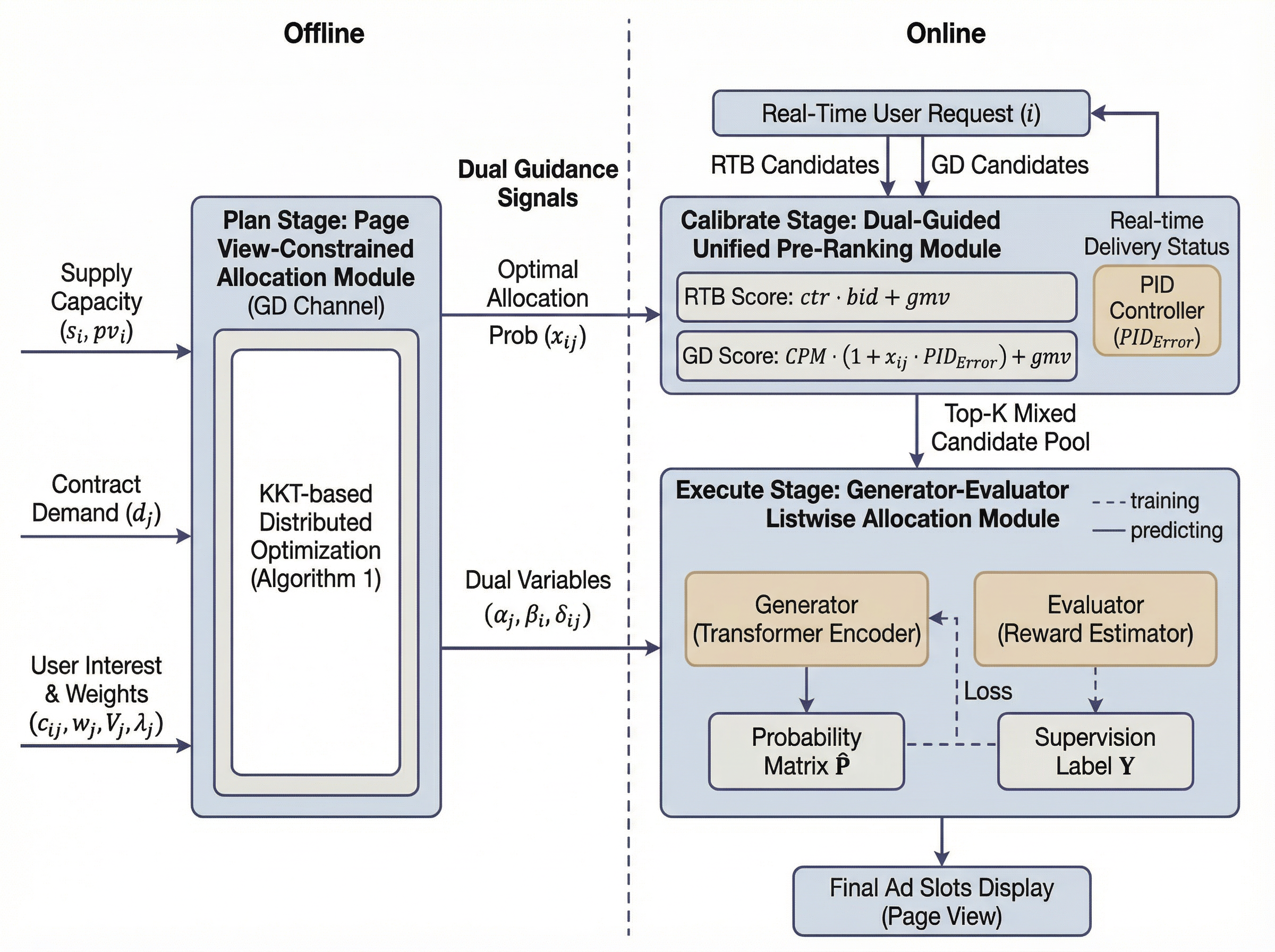}
    \caption{Overview of HMAF.}
    \label{figure: overview}
\end{figure*}

\section{Related Work}

\subsection{Guaranteed Delivery Allocation}

Several approaches focus on the allocation of display ads within GD contexts. Vee et al.~\cite{vee2010optimal} address online allocation using future arrival forecasts, ensuring near-optimal solutions even in adversarial settings. Bharadwaj et al.~\cite{bharadwaj2012shale} propose SHALE, a scalable algorithm for GD advertising, optimizing decisions with minimal state information and linear time complexity. Cheng et al.~\cite{cheng2022adaptive} introduce an adaptive framework that balances contract fulfillment and user interest optimization through real-time pacing adjustments. Fang et al.~\cite{fang2019large} present Xshale, which models GD allocation at the individual user level, incorporating real-time pacing strategies to improve targeting. He et al.~\cite{he2024efficient} propose a local search optimization method to explore feasible allocations while respecting complex multilinear constraints. Li et al.~\cite{li2024bi} introduce a bi-objective optimization framework that balances impressions for new orders and inventory allocation, reducing online delivery risks. More recently, Zhang et al.~\cite{zhang2026beyond} extend GD allocation from the conventional single-slot setting to multi-slot joint optimization by considering page-view constraints.

\subsection{Guaranteed Delivery Optimization}
In addition to allocation, several works focus on optimizing the fulfillment of guaranteed delivery ads. Wu et al.~\cite{wu2021impression} explore auction-based impression allocation, using multi-agent reinforcement learning to optimize allocation decisions and adapt to dynamic traffic patterns. Zhang et al.~\cite{zhang2022control} propose a control-based bidding approach for mobile livestreaming ads, adjusting bidding parameters using feedback control to meet exposure guarantees. Mao et al.~\cite{mao2023end} introduce a unified end-to-end system for inventory prediction and allocation, improving efficiency by integrating forecasting with allocation. Dai et al.~\cite{dai2023fairness} incorporate fairness considerations into GD allocation, balancing contract fulfillment and traffic costs. Wei et al.~\cite{wei2023rltp} present RLTP, a reinforcement learning-based pacing agent that optimizes both guaranteed impression counts and delivery performance through an adaptive reward estimator.

\subsection{GD-RTB Competition}

Combining the GD and RTB models has become a promising approach for maximizing revenue across both streams. Wang et al.~\cite{wang2022conflux} introduce CONFLUX, a unified framework for impression allocation that bridges the GD and RTB markets. By modeling impression allocation at the individual request level, the framework maximizes combined revenue and uses a cascaded learning process to guide a deep model. This model is then transferred to a lightweight model suitable for large-scale online serving. Zhang et al.~\cite{zhang2020request} present RAP, a request-level system for GD advertising that captures fine-grained request attributes and jointly optimizes impression forecasting, sales planning, and supply allocation, ensuring that ads are served effectively. Li et al.~\cite{li2024bi} address the challenge of balancing the allocation of impressions for both new guaranteed contracts and inventory distribution across supply nodes. Their bi-objective optimization framework ensures that impressions are maximized for incoming orders while maintaining a balanced distribution to reduce risks in online serving.


\section{Preliminary}
\subsection{Problem Formulation}
We formulate a multi-slot GD-RTB allocation problem, where GD contracts and RTB ads compete for exposure in a shared stream of multi-slot page view requests. Let $\mathcal{I}$ denote the sequence of incoming user requests (supply), and $\mathcal{J} = \mathcal{J}_{GD} \cup \mathcal{J}_{RTB}$ denote the set of candidate ads (demand). Each request $i \in \mathcal{I}$ may contain multiple ad slots, but for modeling simplicity, we treat $i$ as an atomic supply unit with capacity $s_i$ (e.g., expected impressions).

Our ultimate goal is to maximize the platform's total revenue, which consists of the realized value from RTB auctions and the contractual value from GD fulfillment, subject to strict delivery constraints. Let $x_{ij} \in \{0, 1\}$ and $y_{ij} \in \{0, 1\}$ be binary decision variables indicating whether request $i$ is allocated to a GD contract $j \in \mathcal{J}_{GD}$ or a RTB ad $j \in \mathcal{J}_{RTB}$, respectively. The unified allocation problem is formulated as follows:

\begin{align}
\quad \max_{\mathbf{x}, \mathbf{y}} \quad & \sum_{i \in \mathcal{I}} \left( \sum_{j \in \mathcal{J}_{GD}} \text{CPM}_j \cdot x_{ij} + \sum_{j \in \mathcal{J}_{RTB}} \text{BID}_{ij} \cdot y_{ij} \right) \label{eq:obj} \\
\text{s.t.} \quad & \phi B_j \le \sum_{i \in \Gamma(j)} s_i x_{ij} \le B_j, \quad \forall j \in \mathcal{J}_{GD} \label{eq:gd_constraint} \\
& \sum_{j \in \mathcal{J}_{GD}} x_{ij} + \sum_{j \in \mathcal{J}_{RTB}} y_{ij} \le 1, \quad \forall i \in \mathcal{I} \label{eq:supply_constraint} \\
& s_i (x_{ij} + y_{ij}) \le pv_i, \quad \forall i, j \label{eq:pv_constraint} \\
& x_{ij}, y_{ij} \in \{0, 1\}, \quad \forall i, j \label{eq:binary}
\end{align}
Here, Equation~\ref{eq:obj} represents the total revenue. Constraint~\ref{eq:gd_constraint} enforces the double-sided delivery guarantee for GD contracts, where $B_j$ is the total purchased quantity and $\phi \in [0, 1]$ is the minimum delivery rate required to avoid penalties. Constraint~\ref{eq:supply_constraint} ensures that each supply unit is allocated to at most one ad. 
Constraint~\ref{eq:pv_constraint} represents the fine-grained Page View constraint, preventing over-exposure at specific slot positions.

\subsection{Computational Challenges}

Directly solving the multi-slot GD-RTB allocation problem in an industrial online environment presents two fundamental challenges:

\noindent \textbf{1. NP-Hardness of Combinatorial Allocation}:
The multi-slot GD-RTB allocation problem is computationally intractable, even in an offline setting where all requests $\mathcal{I}$ are known in advance.
Specifically, the combination of the capacity constraint (\ref{eq:supply_constraint}) and the double-sided knapsack-like demand constraint (\ref{eq:gd_constraint}) generalizes the well-known \textit{Multidimensional Knapsack Problem (MKP)}, both of which are NP-hard~\cite{kellerer2004knapsack}. 
The introduction of the Page View constraint (\ref{eq:pv_constraint}) further couples the decision variables across different slots, increasing complexity.

To formally establish NP-hardness, we provide a polynomial-time reduction from MKP to our problem (denoted as $\Pi$). Consider an arbitrary MKP instance defined by:
\begin{itemize}[leftmargin=*]
    \item A set of items $k \in \mathcal{K} = \{1, \dots, N\}$, each with weight $w_k$ and value $v_k$.
    \item A set of knapsacks $m \in \mathcal{M} = \{1, \dots, M\}$, each with capacity $C_m$.
    \item Objective: Maximize $\sum_{m,k} v_k z_{mk}$ subject to $\sum_{k} w_k z_{mk} \le C_m$ and $\sum_{m} z_{mk} \le 1$, where $z_{mk} \in \{0,1\}$.
\end{itemize}

We construct a specific instance of our problem $\Pi$ (defined by Eqs.~\eqref{eq:obj}-\eqref{eq:binary}) using the following mapping $f$:

\textbf{Sets mapping:}
\begin{itemize}[leftmargin=*]
    \item Let the set of impressions $\mathcal{I}$ correspond to the set of items $\mathcal{K}$ (i.e., $|\mathcal{I}| = N$).
    \item Let the set of GD advertisers $\mathcal{J}_{GD}$ correspond to the set of knapsacks $\mathcal{M}$ (i.e., $|\mathcal{J}_{GD}| = M$).
    \item Let the set of RTB advertisers be empty: $\mathcal{J}_{RTB} = \emptyset$.
\end{itemize}

\textbf{Parameter assignment (restriction):}
\begin{itemize}[leftmargin=*]
    \item Set impression consumption $s_i = w_i$.
    \item Set advertiser budget $B_j = C_j$.
    \item Set advertiser value $\text{CPM}_j$ such that the gain matches $v_k$ (a uniform value suffices because even the simplified MKP with item-independent values is NP-hard).
    \item Set the lower bound parameter $\phi = 0$ (relaxing the lower bound constraint).
    \item Set impression capacity $pv_i = 1$ (standard 0/1 constraint).
    \item Since $\mathcal{J}_{RTB} = \emptyset$, all $\text{BID}_{ij}$ terms and variables $y_{ij}$ vanish.
\end{itemize}

Substituting these parameters into our formulation yields:
\begin{enumerate}[leftmargin=*]
    \item Objective: $\max_{\mathbf{x}} \sum_{i \in \mathcal{I}} \sum_{j \in \mathcal{J}_{GD}} \text{CPM}_j x_{ij}$ — matches MKP objective.
    \item GD constraint: Upper bound $\sum_i s_i x_{ij} \le B_j$ (lower bound trivial).
    \item Supply constraint: $\sum_j x_{ij} \le 1$ — matches MKP item assignment constraint.
    \item Binary constraint: $x_{ij} \in \{0,1\}$.
\end{enumerate}

Thus the constructed instance is mathematically identical to MKP. The mapping $f$ involves only constant-time assignments and is clearly polynomial-time. Since MKP is known to be NP-hard~\cite{martello1990knapsack}, it follows that our problem $\Pi$ is NP-hard. Consequently, obtaining an exact integer solution for millions of requests and contracts is impossible within milliseconds.

\noindent \textbf{2. Online Stochasticity:}
In practice, requests $i$ arrive sequentially, and the RTB bids $\text{BID}_{ij}$ are unknown until the auction occurs. This stochastic nature precludes static optimization. While Online Primal-Dual methods can provide asymptotic guarantees, they typically require convex relaxations and cannot directly handle the complex combinatorial constraints (e.g., listwise layout effects) inherent in deep rendering models.

\section{Methodology}


We propose HMAF, a unified multi-slot GD–RTB allocation framework. As illustrated in Figure~\ref{figure: overview}, HMAF consists of three tightly integrated modules:
(1) Plan stage: Page View–Constrained Allocation Module, which formulates GD fulfillment as a page-level constrained optimization problem and derives dual signals to explicitly regulate multi-slot exposure and long-term delivery pressure;
(2) Calibrate stage: Dual-Guided Pre-Ranking Module, which leverages offline dual variables to perform unified candidate filtering and coarse ranking across GD and RTB ads, significantly reducing the decision space while preserving allocation quality; and
(3) Execute stage: Generator–Evaluator Module, which conducts listwise allocation by jointly modeling cross-slot interactions and competitive dependencies to produce the final impression assignment.

\subsection{Page View-Constrained Allocation Module} 

To effectively coordinate ad delivery across multiple slots within the same page view, we first formulate the allocation task as a constrained optimization problem. This formulation captures not only contract fulfillment and user engagement but also fine-grained control over slot-level exposure. In particular, we introduce a novel Page View constraint to limit the number of impressions each slot can serve, thereby preventing over-exposure of high-traffic positions and promoting a more balanced inventory allocation across all ad slots. The detailed objective function and constraints are defined as follows: 
\begin{align}
\arg\min_{x_{ij}} \quad & 
\frac{1}{2} \sum_{j} \sum_{i \in \Gamma(j)} s_i \frac{V_j}{\theta_j} (x_{ij} - \theta_j)^2 
- \sum_{j} w_j \sum_{i \in \Gamma(j)} s_i x_{ij} \\
& - \sum_{j} \lambda_j \sum_{i \in \Gamma(j)} s_i x_{ij} c_{ij} \nonumber \\
\text{s.t.} \quad & 
\sum_{i \in \Gamma(j)} s_i x_{ij} \le d_j, \quad \forall j \\
& \sum_{j \in \Gamma(i)} x_{ij} \le 1, \quad \forall i \\
& x_{ij} \ge 0, \quad \forall i, j\\
& s_ix_{ij} \le pv_i, \quad \forall i, j
\end{align}
where $x_{ij}$ denote the allocation probability from request $i$ (i.e. supply) to contract $j$ (i.e. demand), the user interest between request $i$ and contract $j$ is denoted by $c_{ij}$, and $\lambda_j$ controls the relative importance of user interest for contract $j$ in the objective, each request $i$ has a supply capacity $s_i$, while each contract $j$ requires $d_j$ impressions in total. 
Distinct from prior work, we novelly introduce $pv_i$ to model page-view–level exposure limits for request $i$, which constrains the maximum exposure volume that request $i$ can contribute across multiple ad slots on the same page.  This constraint provides a structural handle for capturing fine-grained supply-side scarcity and enables principled interaction between offline optimization and online learning.
Let $\Gamma(j)$ denote the set of requests that can serve contract $j$, and let $\Gamma(i)$ denote the set of contracts that request $i$ is eligible to serve. Each contract $j$ is further characterized by a priority weight $w_j$, a smoothness parameter $V_j$, and a normalized target delivery ratio $\theta_j = \frac{d_j}{\sum_{i \in \Gamma(j)} s_i}$, where $V_j$ controls the strength of the quadratic fairness regularization that penalizes deviations from $\theta_j$.

\begin{align}
L(\alpha, \beta, \gamma, \delta) = & \frac{1}{2} \sum_{j} \sum_{i \in \Gamma(j)} s_i \frac{V_j}{\theta_j} (x_{ij} - \theta_j)^2 - \sum_{j} w_j \sum_{i \in \Gamma(j)} s_i x_{ij} \nonumber \\
&
- \sum_{j} \lambda_j \sum_{i \in \Gamma(j)} s_i x_{ij} c_{ij} \nonumber + \sum_{j} \alpha_j \left( \sum_{i \in \Gamma(j)} s_i x_{ij} - d_j \right) \\
& + \sum_{i} \beta_i \left( \sum_{j \in \Gamma(i)} x_{ij} - 1 \right) 
- \sum_{j} \sum_{i \in \Gamma(j)} \gamma_{ij} x_{ij} \nonumber \\
& + \sum_{j} \sum_{i \in \Gamma(j)} \delta_{ij} \left( s_i x_{ij} - pv_i \right)
\end{align}

\noindent \textbf{KKT conditions}:
\begin{equation}
s_i \frac{V_j}{\theta_j} (x_{ij} - \theta_j) - w_j s_i - \lambda_j s_i c_{ij} + \alpha_j s_i + \beta_i s_i - \gamma_{ij} + \delta_{ij} s_i = 0
\end{equation}

\begin{equation}
\alpha_j \left( \sum_{i \in \Gamma(j)} s_i x_{ij} - d_j \right) = 0
\label{equ: alpha}
\end{equation}

\begin{equation}
\beta_i \left( \sum_{j \in \Gamma(i)} x_{ij} - 1 \right) = 0
\label{equ: beta}
\end{equation}

\begin{equation}
\gamma_{ij} x_{ij} = 0
\label{equ: gamma}
\end{equation}

\begin{equation}
\delta_{ij} \left( s_i x_{ij} - pv_i \right) = 0
\label{equ: delta}
\end{equation}

\begin{equation}
\alpha_j \geq 0,\quad \beta_i \geq 0,\quad \gamma_{ij} \geq 0,\quad \delta_{ij} \geq 0 
\end{equation}
where $\alpha_j$, $\beta_i$, $\gamma_{ij}$ and $\delta_{ij}$ are the Lagrangian multipliers of constraints
Equation~\ref{equ: alpha}, \ref{equ: beta}, \ref{equ: gamma} and \ref{equ: delta} respectively. According to the KKT conditions, the optimal allocation probability $x_{ij}$ can be derived as:
\begin{equation}
x_{ij} = \max\left\{ 0,\ \theta_j \left(1 + \frac{w_j + \lambda_j c_{ij} - \alpha_j - \beta_i - \delta_{ij}}{V_j} \right) \right\}
\end{equation}

To efficiently optimize the PV-constrained allocation problem, we adopt an online distributed strategy that updates the dual variables $\alpha_j$, $\beta_i$, and $\delta_{ij}$ based on the KKT conditions. These dual variables correspond to the contract demand, request capacity, and slot-level page view constraints, respectively. Instead of solving the primal problem directly, which is computationally expensive in large-scale settings, we leverage the closed-form expression of $x_{ij}$ derived from the KKT optimality condition and iteratively adjust the dual variables using projected gradient steps.

At each iteration, $\alpha_j$ is increased if the total allocated impressions to contract $j$ exceed its demand $d_j$, while $\beta_i$ is updated to ensure the total allocation from request $i$ remains within its capacity. Similarly, $\delta_{ij}$ is adjusted to enforce the fine-grained page view constraint for each ad slot. These updates can be performed in parallel across all $i$ and $j$, making the solution highly scalable and amenable to online deployment. This formulation ensures that the allocation respects all practical delivery constraints while optimizing for efficiency and relevance. The detailed procedure is summarized in Algorithm~\ref{algorithm:test}.




\begin{algorithm}
\caption{PV Constrained Allocation Algorithm}
\label{algorithm:test}
\LinesNumbered
\KwIn {$s_i, d_j, \lambda_j, V_j, w_j, c_{ij}$}
\KwOut {$\alpha_{j}, \beta_{i}, \delta_{ij}, x_{ij}$}
Step 1: Initialize. Set $\alpha_j=w_j+\lambda_jc_{ij}, \theta_j $, calculate $x_{ij}$, and gradient $grad_j$ \\
Step 2:

\For{iteration = 1 to n}{
    update $\alpha_j$ with Equation: $\alpha_j^{t+1} = \alpha_j^t - V_j(1-\frac{d_j(\alpha^t)}{d_j})$ \\
    solve $\beta_i$: $\beta_i \leftarrow \max\left(0, \beta_i + \eta_\beta \left( \sum_{j \in \Gamma(i)} x_{ij} - 1 \right) \right)$ \\
    calculate $\delta_{ij}$: $\delta_{ij} \leftarrow \max\left(0, \delta_{ij} + \eta_\delta \left( s_i x_{ij} - pv_i \right) \right)$ \\
    calculate $x_{ij}$: $x_{ij} = \max\left\{ 0,\ \theta_j \left(1 + \frac{w_j + \lambda_j c_{ij} - \alpha_j - \beta_i - \delta_{ij}}{V_j} \right) \right\}$ \\
    update $grad_j$: $grad_j = \sum_{i \in \Gamma(j)} s_i x_{ij} - d_j$
}
\end{algorithm}

\subsection{Dual-Guided Pre-Ranking Module}

We consider a dual-guided pre-ranking module that jointly serves GD ads and RTB ads within the same ranking list. Let $j$ index candidate ads and $i$ index page view-level requests. For each ad $j$, the online ranking score $rs_j$ is defined as:
\begin{equation}
rs_j =
\begin{cases}
ctr_{ij} \cdot CPCbid_{ij}
+ gmv_j ,
& j \in \{\mathrm{RTB}\}, \\[6pt]
CPM_j \cdot (1 + x_{ij} \cdot PID_{Error}) + gmv_j,
& j \in \{\mathrm{GD}\},
\end{cases}
\end{equation}
Here, we innovatively introduces a dual-guided modulation term, $x_{ij} \cdot PID_{Error}$, where $x_{ij}$ is an allocation confidence score derived from offline optimization described in Algorithm \ref{algorithm:test}, reflecting the competitiveness of contract $j$ to request $i$ within the GD channel under allocation constraints. Unlike traditional PID approaches that apply uniform regulation pressure across all traffic, our method dynamically adjusts the allocation based on context-specific confidence scores, enabling more tailored and responsive ad delivery. The term $PID_{Error}$ represents the real-time delivery pressure of guaranteed ad $j$, which is calculated as follows:

\begin{equation}
PID_{Error} =
\begin{cases}
\mu_0+(1-\mu_0)\frac{dr^{*}-dr_j(t)}{dr^{*}} ,
& dr_j(t) \le dr^{*}, \\[6pt]
\mu_0,
& otherwise,
\end{cases}
\end{equation}
where $dr_j(t) = \frac{\mathcal{N}_j(t)}{\mathcal{B}_j(t)}$ is the delivery rate at time $t$, $\mathcal{N}_j(t)$ is the accumulated impressions achieved up to time $t$, and $\mathcal{B}_j(t)$ denotes the expected number of impressions at time $t$ based on the historical traffic curve. $dr^{*}$ and $\mu_0$ are hyperparameters, serving as the target minimum delivery rate and the baseline competitiveness coefficient, respectively. The parameter $\mu_0$ ensures that the contract retains a minimum boost to prevent abrupt on-off switching, which could destabilize the delivery curve and negatively affect user experience.

\subsection{Generator-Evaluator Listwise Allocation Module}

While the allocation module ensures feasibility, it does not explicitly optimize the relative ordering of candidates across multiple slots.  In multi-slot scenarios, ranking decisions are inherently listwise and position-dependent, which motivates a Generator–Evaluator network that jointly models candidate–position interactions and evaluates ranking lists at the page-view level.

\subsubsection{Dual-Augmented Feature Space}
To bridge the gap between offline page view-constrained allocation in GD channel and online mixed ranking, we explicitly inject the optimization results from Section 4.2 into the Generator-Evaluator network as GD state signals. Unlike traditional user-item features, these signals provide the model with "foresight" regarding system-wide constraints:

GD Guidance Signal ($x_{ij}$): We feed the offline optimal probability $x_{ij}$ directly into the network. This serves as a strong prior, guiding the model to align its listwise generation with the macro-resource plan, ensuring that the final execution does not deviate significantly from the global optimum.

Urgency Signal ($\alpha_j$): The demand shadow price $\alpha_j$ informs the network about fulfillment difficulty. The model implicitly learns to prioritize contracts with lower $\alpha_j$ (indicating high scarcity or under-delivery) and yield traffic for those with high $\alpha_j$ (saturated demand), thereby dynamically balancing pacing.

Competition Anchor ($\beta_j$): The scarcity price $\beta_j$ acts as a valuation baseline for the current slot. It helps the evaluator network gauge whether a candidate ad's utility justifies occupying a premium slot, effectively filtering out low-quality allocations in high-contention contexts.

Topology Constraint Signal ($\delta_{ij}$): By incorporating the position cost $\delta_{ij}$, the model acquires layout-awareness. This signal discourages the generator from placing an ad in a specific position where it has already reached its exposure limit (i.e., high $\delta_{ij}$), promoting diverse and compliant ad placements across the list.

\subsubsection{Generator}
The Generator models the page-view level assignment between candidate ads $\mathbf{X}=\{x_1, \ldots, x_n\}$ and positions $\mathbf{P}=\{p_1, \ldots, p_m\}$ jointly. Leveraging a Transformer-based encoder to capture candidate--position interactions in a shared latent space, it outputs a probability matrix $\hat{P}=\{\hat{p}_{ij}\}$, where $\hat{p}_{ij}$ represents the normalized probability of assigning candidate $x_j$ to position $p_i$:

\begin{equation}
  \hat{p}_{ij} = \frac{\exp\left(x_j^{\top} p_i\right)}{\sum_{k=1}^{n} \exp\left(x_k^{\top} p_i\right)}
\end{equation}

Based on $\hat{P}$, a candidate ranking list $S_g$ is sampled. To enhance exploration, we further generate a set of perturbed ranking lists $\{ S_1, \ldots, S_l \}$ using heuristic rules based on historical bidding patterns.

\subsubsection{Evaluator}
The Evaluator serves as a listwise supervisor that identifies the optimal ranking structure from the set $S = \{S_g, S_1, \ldots, S_l\}$. It estimates a reward $R(S_k)$ for each list by aggregating predicted metrics and delivery constraints to compute the list-level eCPM. The optimal list is selected as $S^\star = \arg\max_{S_k \in S} R(S_k)$ and converted into a binary label matrix $\mathbf{Y} \in \{0,1\}^{m \times n}$, where $Y_{ij}=1$ indicates the assignment of $x_j$ to $p_i$ in $S^\star$. This provides a unified supervision signal balancing revenue efficiency and system stability.

\subsubsection{Training Objective}
The model is optimized via a loss function that combines supervised alignment and reward-aware exploration. Specifically, $\mathcal{L}_{ce}$ aligns the predicted probability $\hat{P}$ with the evaluator-selected optimal assignment $\mathbf{Y}$, while $\mathcal{L}_{re}$ addresses the discrepancy between the model's predicted eCPM ($\hat{e}_{ij}$) and the realized eCPM observed in online logs ($e_{ij}$).
\begin{equation}
    \mathcal{L} = \psi_1 \cdot \underbrace{\left( -\sum_{i,j} Y_{ij}\log \hat{p}_{ij} \right)}_{\mathcal{L}_{ce}} - \psi_2 \cdot \underbrace{\sum_{(i,j) \in \mathcal{O}} \left( \hat{e}_{ij} - e_{ij} \right)}_{\mathcal{L}_{re}}
\end{equation}
where $\mathcal{O}$ represents the set of observed exposure events. By directly penalizing the signed deviation $\hat{e}_{ij} - e_{ij}$, the model is forced to correct systematic over-estimation or under-estimation biases, ensuring that the predicted utility strictly aligns with the real-world settlement data. $\psi_1$ and $\psi_2$ are hyperparameters balancing structural correctness and value calibration.

\section{Experiments}


\subsection{Setup}


\subsubsection{Datasets} 
We collected a large-scale industrial dataset from the Meituan platform's production logs, spanning 32 consecutive days, from 1st January to 1st February, 2026.
The dataset contains 123,546,485 impression samples associated with 193,797 advertisements, of which 9.73\% correspond to GD ads, while the remaining impressions are served through RTB.


\subsubsection{Offline Baselines}
We evaluate the performance of HMAF against a diverse set of representative baselines, including GD-focused models, CF, PID, and AUAF~\cite{cheng2022adaptive}, and GD-RTB models, CONFLUX~\cite{wang2022conflux} and LIBRA~\cite{wang2024follow}:

\begin{itemize}[leftmargin=*]
    \item \textbf{Contract First (CF)}: A heuristic baseline that prioritizes fulfilling GD contracts and allocates residual impressions to RTB, typically optimized via the primal-dual SHALE algorithm~\cite{bharadwaj2012shale}.
    \item \textbf{PID}: A feedback-based method that applies proportional-integral-derivative control to dynamically regulate GD contract delivery and reduce under- or over-delivery.
    \item \textbf{AUAF}~\cite{cheng2022adaptive}: An adaptive unified allocation framework that dynamically balances multiple GD contracts under heterogeneous demand and supply conditions.
    \item \textbf{CONFLUX}~\cite{wang2022conflux}: A request-level fusion framework that unifies GD and RTB allocation via cascaded distillation.
    \item \textbf{LIBRA}~\cite{wang2024follow}: A unified impression allocation approach that enforces fairness across contracts using adversarial reward guidance.
\end{itemize}
To adapt our baselines to our multi-slot setting, we decompose each page view into sequential single-slot decisions, greedily selecting the highest-ranked eligible ad per position.

\subsubsection{Evaluation Metrics}
We consider the following metrics:
\begin{itemize}[leftmargin=*]
    \item \textbf{Group AUC (GA) }: Measures the ranking accuracy of the model within each user session. It serves as a standard metric for verifying that the injection of auxiliary features does not degrade the core CTR estimation capability.
    \item \textbf{Delivery Rate (DR)}: This metric serves as the central measure of GD contract performance and quantifies the fulfillment rate. Since there are multiple contracts in the system, we calculate the weighted delivery rate, $\sum_{j \in \mathcal{J}}weight_j\frac{\sum_{i}x_{ij}}{d_j}$, where $weight_j = \frac{d_j}{\sum_{j \in \mathcal{J}}d_{j}}$.
    \item \textbf{GD Quality (GQ)}: This metric assesses the fulfillment rate of impression quality constraints, specifically in terms of clicks, for GD contracts. We calculate the weighted fullfillment rate as GD quality, $\sum_{j \in \mathcal{J}}weight_j\frac{\sum_{i}x_{ij} \cdot pCTR_{ij}}{click_j}$, where $weight_j = \frac{click_j}{\sum_{j \in \mathcal{J}}click_{j}}$.
    \item \textbf{RTB eCPM (RE)}: As the unified allocation framework releases more high-quality impressions to the RTB market, we calculate the average eCPM (effective cost per mille) of the RTB market. For privacy concerns related to business data, we normalize this metric using the highest eCPM among the compared methods.
    \item \textbf{Utility Rate (UR)}: This metric represents the total utility return from both GD and RTB markets. It considers the sum of GD and RTB incomes while accounting for under delivery penalties. As user traffic and bid landscape evolve over time, we normalize it by the theoretically optimal utility solved using evaluator in Section 4.3. This normalization provides an intuitive and effective metric to evaluate the method’s impact on revenue lift.
\end{itemize}

\subsubsection{Experomental Settings}
During training, the first 31 days of the dataset are used for training, while the last day is reserved for testing, ensuring a robust evaluation of the model's performance on unseen online traffic conditions.All the experiments involving deep learning frameworks are executed on two NVIDIA A100 GPUs. For more detailed hyperparameters, please refer to Table~\ref{tb: hyperparameters}.

\begin{table}[H]
\caption{The detailed hyperparameters of HMAF}
\label{tb: hyperparameters}
\scalebox{0.9}{
\begin{tabular}{ll}
\toprule
Hyperparameters     & Value  \\ 
\midrule
Batch size          & 512    \\
Learning rate       & 3e-4   \\
Optimizer           & LazyAdam  \\

Embedding dim         & 8    \\
Activation function & Swish   \\
$dr*$               & 0.9   \\
$\mu_o$               & 0.1   \\
$\phi_1 $  & 1      \\ 
$\phi_2 $  & -0.2      \\ 

\bottomrule
\end{tabular}}
\end{table}

\subsection{Performance Analysis}

The experimental results in Table~\ref{tab:offline_results} demonstrate that HMAF consistently outperforms all baseline methods across most evaluation metrics, achieving the highest scores in Group AUC (GA), GD Quality (GQ), RTB eCPM (RE), and Utility Rate (UR).

Among GD-focused methods, CF achieves the highest Delivery Rate (DR), which is expected given its heuristic strategy that strictly prioritizes GD contracts before allocating remaining traffic to RTB ads. However, this prioritization comes at the cost of reduced flexibility in revenue optimization. Notably, HMAF attains a highly competitive DR with only a marginal gap from CF, despite not enforcing explicit GD-first rules. This indicates that HMAF can effectively satisfy GD delivery requirements while preserving allocation flexibility for revenue-aware decision-making. PID and AUAF improve GD delivery stability by introducing feedback control or adaptive heuristics, yet they remain primarily optimized for GD contracts and rely on decoupled or rule-based mechanisms when handling RTB ads. As a result, these methods struggle to achieve a balanced trade-off between long-term contract delivery and short-term revenue maximization in a unified allocation setting.

Among GD–RTB baselines, CONFLUX attempts to bridge the two markets through request-level unification, but is designed for single-slot allocation, which limits its ability to capture cross-slot interactions and page-level coordination in multi-slot traffic. LIBRA represents the strongest GD-RTB baseline by incorporating adversarial reward guidance for unified impression allocation. Nevertheless, its decision process lacks explicit mechanisms to dynamically integrate offline constraint signals with online allocation decisions.

In contrast, HMAF emphasizes page-level constraint modeling and coordinated listwise allocation, which jointly enable effective multi-slot decision-making across GD and RTB ads. These design choices allow HMAF to achieve superior overall performance without relying on rigid prioritization rules or heavy online control, leading to consistent gains across both delivery and revenue metrics. 
To verify that these improvements are statistically reliable, we further conduct significance testing against the strongest baselines. The results show that HMAF achieves significant gains on all major metrics, while the reported standard deviations and confidence intervals over multiple runs demonstrate the stability of its performance.

\begin{table}[t]
\centering
\caption{Offline performance comparison, categorized by model type as either GD-focused models (\ding{55}) or GD-RTB models (\ding{51}). The highest score is highlighted in bold, while the second-best result among the baselines is underlined. To ensure robustness, all metrics are reported as the mean and standard deviation over ten independent runs.} 
\label{tab:offline_results}
\resizebox{0.48\textwidth}{!}{
\begin{tabular}{clccccc}
\toprule
\textbf{Type} & \textbf{Model} & \textbf{GA} & \textbf{DR} & \textbf{GQ} & \textbf{RE} & \textbf{UR} \\
\midrule
\ding{55} & CF 
   & \makecell{0.7203 \\ \scriptsize{$(\pm 0.0032)$}}
   & \makecell{\textbf{0.9680} \\ \scriptsize{$(\pm 0.0000)$}}
   & \makecell{0.8351 \\ \scriptsize{$(\pm 0.0041)$}}
   & \makecell{0.7333 \\ \scriptsize{$(\pm 0.0050)$}}
   & \makecell{0.8273 \\ \scriptsize{$(\pm 0.0043)$}} \\
\ding{55} & PID 
   & \makecell{0.7839 \\ \scriptsize{$(\pm 0.0027)$}}
   & \makecell{0.9273 \\ \scriptsize{$(\pm 0.0025)$}}
   & \makecell{0.8604 \\ \scriptsize{$(\pm 0.0033)$}}
   & \makecell{0.7734 \\ \scriptsize{$(\pm 0.0038)$}}
   & \makecell{0.9379 \\ \scriptsize{$(\pm 0.0029)$}} \\
\ding{55} & AUAF 
   & \makecell{0.8026 \\ \scriptsize{$(\pm 0.0024)$}}
   & \makecell{0.9408 \\ \scriptsize{$(\pm 0.0022)$}}
   & \makecell{0.8948 \\ \scriptsize{$(\pm 0.0028)$}}
   & \makecell{0.8023 \\ \scriptsize{$(\pm 0.0032)$}}
   & \makecell{0.8953 \\ \scriptsize{$(\pm 0.0030)$}} \\
\midrule
\ding{51} & CONFLUX 
   & \makecell{0.8274 \\ \scriptsize{$(\pm 0.0025)$}}
   & \makecell{0.9493 \\ \scriptsize{$(\pm 0.0021)$}}
   & \makecell{0.9264 \\ \scriptsize{$(\pm 0.0029)$}}
   & \makecell{0.8673 \\ \scriptsize{$(\pm 0.0035)$}}
   & \makecell{0.9308 \\ \scriptsize{$(\pm 0.0028)$}} \\
\ding{51} & LIBRA 
   & \makecell{\underline{0.8538} \\ \scriptsize{$(\pm 0.0031)$}}
   & \makecell{0.9554 \\ \scriptsize{$(\pm 0.0024)$}}
   & \makecell{\underline{0.9450} \\ \scriptsize{$(\pm 0.0035)$}}
   & \makecell{\underline{0.9132} \\ \scriptsize{$(\pm 0.0042)$}}
   & \makecell{\underline{0.9513} \\ \scriptsize{$(\pm 0.0033)$}} \\
\midrule
\ding{51} & \textbf{HMAF} 
   & \makecell{\textbf{0.8984} \\ \scriptsize{$(\pm 0.0021)$}}
   & \makecell{\underline{0.9638} \\ \scriptsize{$(\pm 0.0018)$}}
   & \makecell{\textbf{0.9751} \\ \scriptsize{$(\pm 0.0019)$}}
   & \makecell{\textbf{1.0000} \\ \scriptsize{$(\pm 0.0000)$}}
   & \makecell{\textbf{0.9812} \\ \scriptsize{$(\pm 0.0017)$}} \\
\bottomrule
\end{tabular}
}
\end{table}

\subsection{Ablation Study}

T
Table~\ref{tab: ablation} summarizes the ablation results of HMAF. Overall, removing any component results in noticeable performance degradation. In particular, excluding the PV module leads to a significant drop in allocation accuracy.
This decline occurs because the PV module explicitly captures page-level exposure constraints and propagates the corresponding dual information to the online decision stage. Without this guidance, the model tends to make slot-wise decisions independently, which weakens its ability to coordinate impressions across multiple slots on the same page. Removing the GE module also hurts all metrics, showing that listwise allocation is necessary for balancing GD delivery and RTB revenue in the multi-slot setting. In contrast, removing the PR module yields slightly higher DR and UR scores. This effect can be attributed to PR module’s candidate pruning mechanism, which trades off a small amount of allocation flexibility for efficiency by reducing the candidate space. Notably, PR reduces the end-to-end optimization time from approximately two days to 15 minutes, while introducing only marginal differences in allocation performance. These results demonstrate that PR provides an effective accuracy–efficiency trade-off and is essential for enabling scalable deployment in large-scale production systems.

\begin{table}[t]
\centering
\caption{Ablation study performances by removing Page View-Constrained Allocation Module (PV), Dual-Guided Pre-Ranking Module (PR) and Generator-Evaluator Listwise Allocation Module (GE)}
\label{tab: ablation}
\begin{tabular}{lccccc}
\toprule
\textbf{Model} & \textbf{GA} & \textbf{DR} & \textbf{GQ} & \textbf{RE} & \textbf{UR} \\
\midrule
w/o PV & 0.8419 & 0.9458 & 0.9503 & 0.9496 & 0.9647 \\
w/o PR & 0.7943 & 0.9647 & 0.9604 & 0.9672 & 0.9810 \\
w/o GE & 0.8294 & 0.9583 & 0.9382 & 0.9252 & 0.9458 \\
\midrule
\textbf{HMAF} & \textbf{0.8984} & \textbf{0.9638} & \textbf{0.9751} & \textbf{1.0000} & \textbf{0.9812}\\
\bottomrule
\end{tabular}
\end{table}

\begin{figure}[t]
    \centering
    \begin{minipage}[t]{0.238\textwidth}
        \centering
        \includegraphics[width=1.0\linewidth]{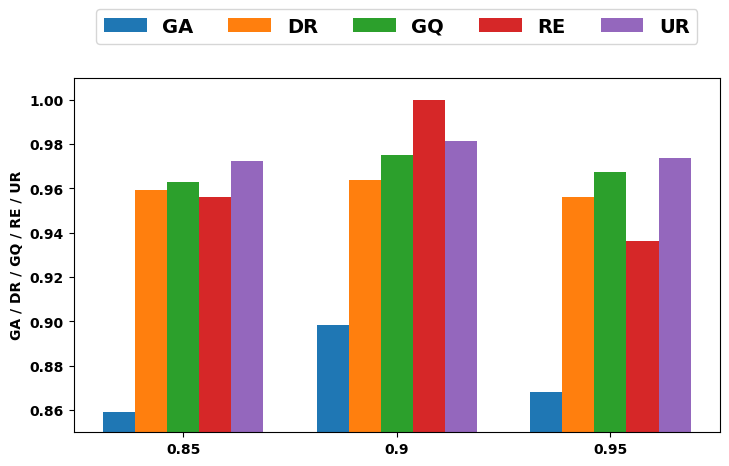}
        \caption*{(a) Parameter analysis on $dr^{*}$.}
    \end{minipage}\hfill
    \begin{minipage}[t]{0.238\textwidth}
        \centering
        \includegraphics[width=1.0\linewidth]{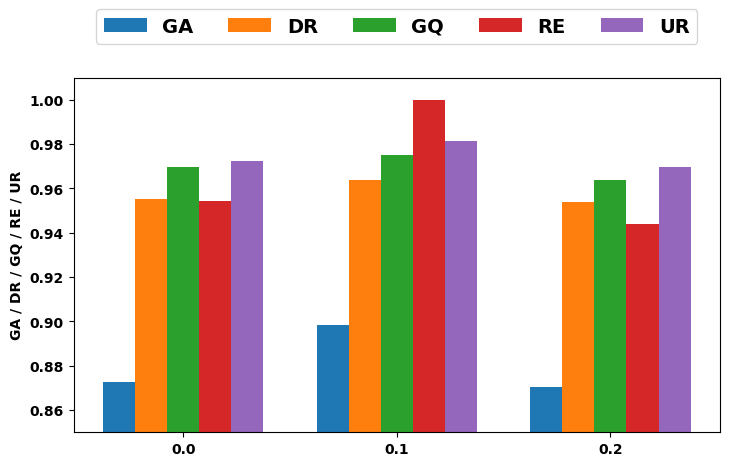}
        \caption*{(b) Parameter analysis on $\mu_0$.}
    \end{minipage}

    \vspace{0.6em}

    \begin{minipage}[t]{0.48\textwidth}
        \centering
        \includegraphics[width=1.0\linewidth]{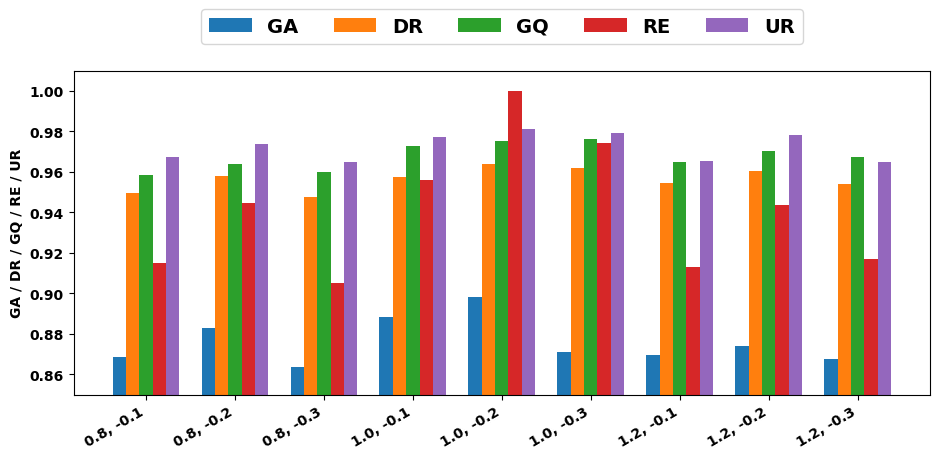}
        \caption*{(c) Joint parameter analysis on $(\psi_1,\psi_2)$.}
    \end{minipage}

    \caption{Parameter sensitivity analysis of HMAF.}
    \label{fig: param}
\end{figure}

\subsection{Parameters Analysis}
We evaluated the impact of four key hyperparameters: $dr^{*}$ sets the target minimum delivery rate for GD contracts, ensuring that the system meets the required delivery levels; $\mu_0$ acts as a baseline competitiveness coefficient, influencing the allocation of resources between GD and RTB ads;  $\psi_1$ and $\psi_2$ control the relative importance of the loss terms, with $\psi_1$ balancing the importance of the page-view constrained allocation and $\psi_2$ managing the trade-off between real-time revenue and long-term contract fulfillment.

As shown in Figure~\ref{fig: param}, the selected parameter configuration achieves the best overall performance. If $dr^{*}$ is too low, the model may prioritize contract fulfillment at the expense of maximizing revenue, leading to suboptimal performance in the RTB market. On the other hand, if $dr^{*}$ is set too high, the system might overemphasize fulfillment, which could result in missed revenue opportunities. Similarly, $\mu_0$ influences the trade-off between ensuring delivery and maximizing RTB revenue. A $\mu_0$ that is too low reduces the system's ability to prioritize GD contracts, leading to potential delivery issues. Conversely, an excessively high $\mu_0$ results in an overemphasis on GD contracts, reducing flexibility in RTB allocation and limiting the overall revenue potential. In our joint analysis of $\psi_1$ and $\psi_2$, we observe that when these parameters are set too low, the system struggles to effectively balance the dual objectives of contract fulfillment and revenue maximization. On the other hand, setting them too high results in overfitting, where the system either excessively prioritizes GD contracts or diminishes RTB revenue generation. The optimal performance is achieved when these parameters are appropriately balanced, enabling the model to dynamically adapt to varying traffic conditions while ensuring efficient allocation across multiple slots.

\section{Online A/B Testing}

\subsection{Setup}
The experiments were conducted on the Meituan advertising platform with live traffic. We employed a rigorous A/B testing framework:

\begin{itemize}[leftmargin=*]
    \item \textbf{Control Group:} The legacy system using a standard PID-based pacing algorithm with a greedy ranking strategy (RTB-first).
    \item \textbf{Treatment Group:} Our proposed HMAF framework, including the Offline Primal-Dual Solver, Pacing-Adaptive Pre-Ranking, and Dual-Augmented GE Reranking.
    \item \textbf{Traffic Split:} Traffic was randomly split by user ID buckets (50\% vs. 50\%) to ensure fair comparison. The experiment lasted for 2 months, which is reasonably representative as it covers two complete business cycles with both weekday and weekend traffic patterns.
\end{itemize}

\subsection{System Efficiency and Stability}
In production, the offline stage runs on 20 machines, each equipped with 32-core Intel Xeon Gold 6130 CPUs, and processes approximately 100 million SKUs in around two hours. The online serving stage is lightweight, pre-ranking takes less than 5 ms per ad, and GE inference remains within 50 ms per page view. The daily infrastructure cost is approximately \$150.

To improve robustness under traffic shifts, HMAF adopts two daily update mechanisms: (1) \textit{Daily Re-planning}, the offline KKT optimization is re-run every day using the latest traffic forecasts, and  (2) \textit{Strict Gating}, the GE network is retrained daily and deployed only when its offline Group-AUC fluctuation is within 5\% of the serving model.

For newly created GD contracts with insufficient real-time delivery data, HMAF uses a cold-start strategy based on historical [City] $\times$ [Category] averages to initialize the dual variables. This provides a reasonable exposure baseline before enough online data is accumulated, after which the PID controller performs minor adjustments around the globally coordinated allocation.

\subsection{Online A/B Test Results}
We adopt a comprehensive set of metrics covering three dimensions:
\begin{itemize}[leftmargin=*]
    \item \textbf{Platform Revenue:} Measured by \textit{Revenue Per Mille (RPM)}, eCPM and total advertising revenue (TR).
    \item \textbf{Contract Delivery:} Measured by \textit{Delivery Rate (DR)}—the ratio of delivered impressions to the guaranteed quantity.
    \item \textbf{Merchant Efficiency:} Measured by \textit{ROI (Return on Investment)}, ensuring that increased revenue does not come at the cost of advertiser satisfaction.
\end{itemize}


Our framework achieves a significant \textit{Pareto Improvement}. The delivery rate increased by \textbf{3.72\%}, effectively eliminating under-delivery issues. Simultaneously, total advertisement revenue increased by \textbf{1.59\%}. This counter-intuitive result (usually better fulfillment hurts revenue) is attributed to the fine-grained optimization of our GE model: by utilizing the $\delta_{ij}$ (slot constraint) and $\beta_i$ (scarcity) signals, the model intelligently shifts GD impressions to non-premium slots or low-competition traffic, releasing high-value inventory for RTB ads. 
We further verify that this joint improvement is statistically reliable. Over the two-month online experiment, we performed a paired t-test on the daily delivery-rate differences between the treatment and control groups, yielding a p-value of 0.018 (<0.05). The 95\% confidence interval for the delivery-rate lift is [3.04\%, 4.27\%]. For total advertisement revenue, the p-value is 0.024, with a 95\% confidence interval of [1.23\%, 2.15\%] for the revenue lift.





\section{Conclusion}

In this paper, we propose HMAF, a hierarchical multi-slot GD-RTB allocation framework that effectively balances long-term contract fulfillment with short-term revenue maximization in GD-RTB advertising systems. HMAF introduces a novel \emph{Plan–Calibrate–Execute} paradigm, which tightly integrates offline constraint optimization with online neural allocation, enabling systematic coordination across the allocation process. This paradigm ensures both precise global planning and dynamic real-time adjustments, making it well-suited for the complex challenges of GD–RTB systems. By introducing page view constraints and deriving dual signals that capture global scarcity and delivery pressure, HMAF enhances coordination across GD and RTB ads in multi-slot environments. Experimental results demonstrate that HMAF significantly outperforms baseline models, achieving higher revenue and improved contract delivery rates. 
HMAF has also been deployed in Meituan’s production advertising system, where it brings measurable gains under large-scale live traffic. Since its core formulation only relies on guaranteed delivery requirements, real-time bidding values, and multi-slot exposure constraints, the framework can be adapted to other advertising platforms with similar GD--RTB allocation settings.



\bibliographystyle{ACM-Reference-Format}
\bibliography{sample-base}

@inproceedings{ghosh2009bidding,
  title={Bidding for representative allocations for display advertising},
  author={Ghosh, Arpita and McAfee, Preston and Papineni, Kishore and Vassilvitskii, Sergei},
  booktitle={International workshop on internet and network economics},
  pages={208--219},
  year={2009},
  organization={Springer}
}

@inproceedings{dai2023fairness,
  title={Fairness-aware guaranteed display advertising allocation under traffic cost constraint},
  author={Dai, Liang and Zu, Zhonglin and Wu, Hao and Wang, Liang and Zheng, Bo},
  booktitle={Proceedings of the ACM Web Conference 2023},
  pages={3572--3580},
  year={2023}
}

@inproceedings{wang2022conflux,
  title={CONFLUX: A Request-level Fusion Framework for Impression Allocation via Cascade Distillation},
  author={Wang, XiaoYu and Tan, Bin and Guo, Yonghui and Yang, Tao and Huang, Dongbo and Xu, Lan and Freris, Nikolaos M and Zhou, Hao and Li, Xiang-Yang},
  booktitle={Proceedings of the 28th ACM SIGKDD Conference on Knowledge Discovery and Data Mining},
  pages={4070--4078},
  year={2022}
}

@inproceedings{he2024efficient,
  title={An Efficient Local Search Algorithm for Large GD Advertising Inventory Allocation with Multilinear Constraints},
  author={He, Xiang and Mao, Wuyang and Xu, Zhenghang and Gu, Yuanzhe and Huang, Yundu and Zu, Zhonglin and Wang, Liang and Zhao, Mengyu and Zou, Mengchuan},
  booktitle={Proceedings of the 30th ACM SIGKDD Conference on Knowledge Discovery and Data Mining},
  pages={1040--1049},
  year={2024}
}

@inproceedings{vee2010optimal,
    title={Optimal online assignment with forecasts}, author={Vee, Erik and Vassilvitskii, Sergei and Shanmugasundaram, Jayavel},
    booktitle={Proceedings of the 11th ACM conference on Electronic commerce},
    pages={109--118},
    year={2010}
}

@inproceedings{bharadwaj2012shale,
    title={Shale: an efficient algorithm for allocation of guaranteed display advertising},   
    author={Bharadwaj, Vijay and Chen, Peiji and Ma, Wenjing and Nagarajan, Chandrashekhar and Tomlin, John and Vassilvitskii, Sergei and Vee, Erik and Yang, Jian},
    booktitle={Proceedings of the 18th ACM SIGKDD international conference on Knowledge discovery and data mining}, 
    pages={1195--1203},
    year={2012}
}

@inproceedings{cheng2022adaptive,   
    title={An adaptive unified allocation framework for guaranteed display advertising},   
    author={Cheng, Xiao and Liu, Chuanren and Dai, Liang and Zhang, Peng and Fang, Zhen and Zu, Zhonglin},   
    booktitle={Proceedings of the Fifteenth ACM International Conference on Web Search and Data Mining},   
    pages={132--140},   
    year={2022} 
}

@inproceedings{fang2019large,   
    title={Large-scale personalized delivery for guaranteed display advertising with real-time pacing},   
    author={Fang, Zhen and Li, Yang and Liu, Chuanren and Zhu, Wenxiang and Zheng, Yu and Zhou, Wenjun},   
    booktitle={2019 IEEE International Conference on Data Mining (ICDM)},   
    pages={190--199},   
    year={2019},   
    organization={IEEE} 
}

@inproceedings{zhang2020request,   
    title={A request-level guaranteed delivery advertising planning: Forecasting and allocation},   
    author={Zhang, Hong and Zhang, Lan and Xu, Lan and Ma, Xiaoyang and Wu, Zhengtao and Tang, Cong and Xu, Wei and Yang, Yiguo},   
    booktitle={Proceedings of the 26th ACM SIGKDD International Conference on Knowledge Discovery \& Data Mining},   
    pages={2980--2988},   
    year={2020} 
}

@inproceedings{mao2023end,   
    title={End-to-end inventory prediction and contract allocation for guaranteed delivery advertising},   
    author={Mao, Wuyang and Liu, Chuanren and Huang, Yundu and Zu, Zhonglin and Harshvardhan, M and Wang, Liang and Zheng, Bo},   
    booktitle={Proceedings of the 29th ACM SIGKDD Conference on Knowledge Discovery and Data Mining},   
    pages={1677--1686},   
    year={2023} 
}

@inproceedings{li2024bi,   
    title={Bi-Objective Contract Allocation for Guaranteed Delivery Advertising},   
    author={Li, Yan and Huang, Yundu and Mao, Wuyang and Ye, Furong and He, Xiang and Zu, Zhonglin and Cai, Shaowei},   
    booktitle={Proceedings of the 30th ACM SIGKDD Conference on Knowledge Discovery and Data Mining},   
    pages={1691--1700},   
    year={2024} 
}

@inproceedings{wu2021impression,   
    title={Impression allocation and policy search in display advertising},   
    author={Wu, Di and Chen, Cheng and Chen, Xiujun and Pan, Junwei and Yang, Xun and Tan, Qing and Xu, Jian and Lee, Kuang-Chih},   
    booktitle={2021 IEEE International Conference on Data Mining (ICDM)},   
    pages={749--756},      
    year={2021},   
    organization={IEEE} 
}

@inproceedings{zhang2022control,   
    title={Control-based Bidding for Mobile Livestreaming Ads with Exposure Guarantee},   
    author={Zhang, Haoqi and Jin, Junqi and Zheng, Zhenzhe and Wu, Fan and Xu, Haiyang and Xu, Jian},   
    booktitle={Proceedings of the 31st ACM International Conference on Information \& Knowledge Management},   
    pages={2539--2548},   
    year={2022} 
}

@inproceedings{bhalgat2012online,   
    title={Online allocation of display ads with smooth delivery},   
    author={Bhalgat, Anand and Feldman, Jon and Mirrokni, Vahab},   
    booktitle={Proceedings of the 18th ACM SIGKDD international conference on Knowledge discovery and data mining},   
    pages={1213--1221},   
    year={2012} 
}

@inproceedings{zhang2016feedback,   
    title={Feedback control of real-time display advertising},   
    author={Zhang, Weinan and Rong, Yifei and Wang, Jun and Zhu, Tianchi and Wang, Xiaofan},   
    booktitle={Proceedings of the Ninth ACM International Conference on Web Search and Data Mining},   
    pages={407--416},   
    year={2016} 
}

@inproceedings{wei2023rltp,   
    title={RLTP: Reinforcement Learning to Pace for Delayed Impression Modeling in Preloaded Ads},   
    author={Wei, Penghui and Chen, Yongqiang and Liu, Shaoguo and Wang, Liang and Zheng, Bo},   
    booktitle={Proceedings of the 29th ACM SIGKDD Conference on Knowledge Discovery and Data Mining},   
    pages={5204--5214},   
    year={2023} 
}

@inproceedings{chen2011real,   
    title={Real-time bidding algorithms for performance-based display ad allocation},   
    author={Chen, Ye and Berkhin, Pavel and Anderson, Bo and Devanur, Nikhil R},   
    booktitle={Proceedings of the 17th ACM SIGKDD international conference on Knowledge discovery and data mining},   
    pages={1307--1315},   
    year={2011} 
}

@inproceedings{hojjat2014delivering,
  title={Delivering guaranteed display ads under reach and frequency requirements},
  author={Hojjat, Ali and Turner, John and Cetintas, Suleyman and Yang, Jian},
  booktitle={Proceedings of the AAAI Conference on Artificial Intelligence},
  volume={28},
  number={1},
  year={2014}
}

@inproceedings{wang2015real,
  title={Real-time bidding: A new frontier of computational advertising research},
  author={Wang, Jun and Yuan, Shuai},
  booktitle={Proceedings of the Eighth ACM International Conference on Web Search and Data Mining},
  pages={415--416},
  year={2015}
}

@inproceedings{yuan2013real,
  title={Real-time bidding for online advertising: measurement and analysis},
  author={Yuan, Shuai and Wang, Jun and Zhao, Xiaoxue},
  booktitle={Proceedings of the seventh international workshop on data mining for online advertising},
  pages={1--8},
  year={2013}
}

@inproceedings{zhang2014optimal,
  title={Optimal real-time bidding for display advertising},
  author={Zhang, Weinan and Yuan, Shuai and Wang, Jun},
  booktitle={Proceedings of the 20th ACM SIGKDD international conference on Knowledge discovery and data mining},
  pages={1077--1086},
  year={2014}
}

@inproceedings{wang2024follow,
  title={Follow the libra: guiding fair policy for unified impression allocation via adversarial rewarding},
  author={Wang, Xiaoyu and Guo, Yonghui and Tan, Bin and Yang, Tao and Huang, Dongbo and Xu, Lan and Zhou, Hao and Li, Xiangyang},
  booktitle={Proceedings of the 17th ACM International Conference on Web Search and Data Mining},
  pages={750--759},
  year={2024}
}

@book{kellerer2004knapsack,
  title={Knapsack Problems},
  author={Kellerer, Hans and Pferschy, Ulrich and Pisinger, David},
  year={2004},
  publisher={Springer},
  address={Berlin, Heidelberg}
}

@book{martello1990knapsack,
  title={Knapsack problems: algorithms and computer implementations},
  author={Martello, Silvano and Toth, Paolo},
  year={1990},
  publisher={John Wiley \& Sons, Inc.}
}

@article{zhang2026beyond,
  title={Beyond Single Slot: Joint Optimization for Multi-Slot Guaranteed Display Advertising},
  author={Zhang, Zhaoqi and Deng, Jiaming and Xie, Miao and Cai, Linyou and Xie, Qianlong and Wang, Xingxing and Luo, Siqiang and Cong, Gao},
  journal={arXiv preprint arXiv:2605.21556},
  year={2026}
}

\end{document}